\documentclass[review]{elsarticle}

\usepackage{amsfonts, mathrsfs, amsmath}
\usepackage{lineno,hyperref}
\usepackage{cleveref}
\usepackage{graphicx}
\modulolinenumbers[5]

\journal{Topology and its applications}

\newcommand{\X}{\mathbb{X}}
\newcommand{\Y}{\mathbb{Y}}
\newcommand{\R}{\mathbb{R}}
\newcommand{\N}{\mathbb{N}}
\newcommand{\T}{\mathfrak{T}}

\begin{document}

\begin{frontmatter}

\title{Investigation of Flash Crash via Topological Data Analysis}


\author[mymainaddress]{Wonse Kim\fnref{fn1,fn2}}
\fntext[fn1]{W. Kim and G. Lee are partially supported by BK21 PLUS SNU Mathematical Sciences Division.}
\fntext[fn2]{W. Kim and W. Kook are partially supported by National Research Foundation of Korea (NRF) Grant funded by the Korean Government (MSIP) [No.\ 2018R1A2A3075511].}
\ead{aquinasws@snu.ac.kr}

\author[mymainaddress]{Younng-Jin Kim\fnref{fn3}}
\fntext[fn3]{Y.-J. Kim is supported by National Research Foundation of Korea (NRF) Grant funded by the Korean Government (NRF-2015-Global Ph.D.\ Fellowship Program).}
\ead{sptz@snu.ac.kr}

\author[mymainaddress]{Gihyun Lee\fnref{fn1}}
\ead{gihyun.math@gmail.com}

\author[mymainaddress]{Woong Kook\corref{mycorrespondingauthor}\fnref{fn2,fn4}}
\cortext[mycorrespondingauthor]{Corresponding author}
\fntext[fn4]{W. Kook is partially supported by National Research Foundation of Korea (NRF) Grant funded by the Korean Government (MSIP) [No.\ 2017R1A5A1015626].}
\ead{woongkook@snu.ac.kr}

\address[mymainaddress]{Department of Mathematical Sciences, Seoul National University, Seoul, South Korea}

\begin{abstract}
Topological data analysis has been acknowledged as one of the most successful mathematical data analytic methodologies in various fields including medicine, genetics, and image analysis. In this paper, we explore the potential of this methodology in finance by applying persistence landscape and dynamic time series analysis to analyze an extreme event in the stock market, known as \emph{Flash Crash}. We will provide results of our empirical investigation to confirm the effectiveness of our new method not only for the characterization of this extreme event but also for its prediction purposes.
\end{abstract}


\end{frontmatter}

\section{Introduction}
Topological data analysis (TDA) is a relatively new, large, and growing field in mathematics using a wide variety of techniques based on homology theory and statistics. TDA extracts hidden intelligence from the \emph{shape} of data that previous data analytic methods could not reveal, and various fields including medicine, image analysis, and genetics ~\cite{ref7, ref9} have tremendously benefited from this intriguing methodology. Nevertheless, in the field of finance, there has been only a small number of applications so far~\cite{ref5}. The purpose of this paper is to apply persistence landscape to detect extreme anomalies in financial market, and demonstrate how TDA can provide not only global characterization of data but also dynamic local features for prediction purposes.

In May 6, 2010, there was an unprecedented sudden intraday stock market crash in U.S. financial markets known as \emph{Flash Crash}, and that event has recently emerged as an interesting research topic in finance~\cite{ref3, ref4, ref6, ref8}. In this paper, we combine the methods of TDA including persistent landscape proposed by~\cite{ref5} and techniques of time series analysis, and create a new method for detecting intraday stock market crashes based on $L^1$-norm of persistence landscape. We then empirically show that our method can characterize and predict the event of flash crash as well.

The rest of the paper is organized as follows: \Cref{sec:2} introduces background and techniques of topological data analysis which are used in the paper, and provide a summary of the flash crash event. \Cref{sec:3} describes our data constructed from major stock market indexes, and explains our new method based on TDA and statistical analysis. \Cref{sec:4} reports empirical results to validate effectiveness of our methods for prediction purposes. \Cref{sec:5} concludes the paper with further practical implications of our results.

\section{Background} \label{sec:2}
\subsection{Topological Data Analysis}
In this section, we gather key concepts of topological data analysis that will be used later in the paper. 
Consider a data set $\X =\{x_1, x_2, \ldots, x_n \}$, which is a finite subset of a Euclidean space $\R^d$. Let $R(\X ,\epsilon)$ denote the \emph{Vietoris-Rips complex} for the data set $\X$ and a distance $\epsilon>0$, i.e., $R(\X ,\epsilon)$ is the simplicial complex on the vertex set $\X$ such that
\begin{equation}\label{eq:1}
\mbox{a subset } \sigma \mbox{ of } \X \mbox{ is a simplex in } R(\X ,\sigma) \mbox{ if and only if } d(x,y)<\epsilon \mbox{ for all } x,y\in\sigma.
\end{equation}
Here $d(x,y)$ is the Euclidean distance between $x$ and $y$.

In what follows, let us denote by $H_i(K)$ the $i$-th homology group of a simplicial complex $K$. Throughout this paper we only use homology with real coefficients. It follows from the very definition of Vietoris-Rips complex~\eqref{eq:1} that we have $R(\X ,\epsilon)\subset R(\X ,\epsilon')$ whenever $\epsilon<\epsilon'$. Using this we see that the Vietoris-Rips complexes $R(\X ,\epsilon)$, $\epsilon>0$, form a filtration, which is called the \emph{Vietoris-Rips filtration}. This filtration induces homomorphisms between homology groups. That is, for each $i$, we get a canonical linear map,
\begin{equation}\label{eq:2}
H_i (R(\X ,\epsilon))\longrightarrow H_i (R(\X ,\epsilon')) \quad \mbox{ whenever } \quad \epsilon<\epsilon'.
\end{equation}

The set of homology groups $\{ H_i (R(\X ,\epsilon )) \}_{\epsilon >0}$ and the linear maps~\eqref{eq:2} form a \emph{persistence module} in the sense of~\cite{ref2}, which enables us to track the \emph{birth} and \emph{death}  of homology classes as the parameter $\epsilon >0$ increases. More precisely, given $\epsilon >0$ and $i$, let $0\neq \alpha \in H_i (R(\X ,\epsilon ))$. By using the Vietoris-Rips filtration we see that there exist positive real numbers $b_\alpha$ and $d_\alpha$ satisfying $b_\alpha\leq \epsilon \leq d_\alpha$ and

\begin{itemize}
\item $\alpha$ is not the image of the map $H_i(R(\X ,\delta ))\rightarrow H_i(R(\X ,\epsilon ))$ if $\delta <b_\alpha$.
\item For $b_\alpha\leq\delta\leq\epsilon$, there is $0\neq\beta\in H_i (R(\X ,\delta ))$ whose image by the map $H_i(R(\X ,\delta ))\rightarrow H_i(R(\X ,\epsilon ))$ is $\alpha$.
\item For $\epsilon\leq\delta\leq d_\alpha$, the image of $\alpha$ under the map $H_i(R(\X ,\epsilon ))\rightarrow H_i(R(\X ,\delta ))$ is non-zero.
\item The map $H_i(R(\X ,\epsilon ))\rightarrow H_i(R(\X ,\delta ))$ sends $\alpha$ to the zero in $H_i(R(\X ,\delta ))$ if $\delta>d_\alpha$.
\end{itemize}

Let  $(b_\alpha, d_\alpha)$ denote the open interval corresponding to the birth and death of a given homology class $\alpha$. In what follows let $\T_i$ denote the set of open intervals,
\begin{equation*}
\{(b_\alpha ,d_\alpha )  \mid \alpha \mbox{ is a non-zero } i\mbox{-dimensional homology class} \}.
\end{equation*}
Here the elements of $\T_i$ are counted with multiplicity, i.e., if $\alpha$ and $\beta$ are distinct $i$-dimensional homology classes, then $(b_\alpha ,d_\alpha )$ and  $(b_\beta ,d_\beta )$ are regarded as different elements of $\T_i$ even though these two open intervals may coincide.

Given a pair of real numbers $b<d$, set
\begin{equation*}
f_{(b,d)}(x)=
\left\{
\begin{array}{cl}
x-b & \mbox{for }  b\leq x\leq \dfrac{b+d}{2} \\
-x+d & \mbox{for } \dfrac{b+d}{2}\leq x\leq d \\
0 & \mbox{otherwise.}
\end{array}
\right.
\end{equation*}
The \emph{persistence landscape} associated with the data set $\X$ is a function $\lambda(\X):\N\times\R\rightarrow \R$ which encodes the information about the birth and death of homology classes~\cite{ref1}. It is defined by
\begin{equation*}
\lambda(\X)(k,x)=k\text{-max} \{f_{(b,d)}(x)\mid (b,d)\in\T_i \}, \quad (k,x)\in\N\times\R.
\end{equation*}
Here $k$-max denotes the $k$-th largest value counted with multiplicity. Note that the persistence landscape $\lambda(\X)$ can be seen as a sequence of functions $\{\lambda(\X)_k \}_{k\geq 1}$. 

For $1\leq p<\infty$, the $L^p$-norm of the persistence landscape $\lambda(\X)= \{\lambda(\X)_k \}_{k\geq 1}$ is defined by
\begin{equation}\label{eq:3}
\lVert\lambda(\X)\rVert_p=\sum^\infty_{k=1}\left(\int^\infty_{-\infty}\lvert\lambda(\X)_k(x)\rvert^p \mathrm{d}x\right)^{\frac{1}{p}}.
\end{equation}
As $\X$ is a finite set, all simplicial complexes $R(\X ,\epsilon)$, $\epsilon>0$, are finite. This shows that for $k$ large enough, we have $\lambda(\X)_k=0$, and hence there is no issue of convergence of the series given in~\eqref{eq:3}.

\subsection{Flash Crash}
\begin{figure}[h]
\centering
\includegraphics[width=1.0\textwidth]{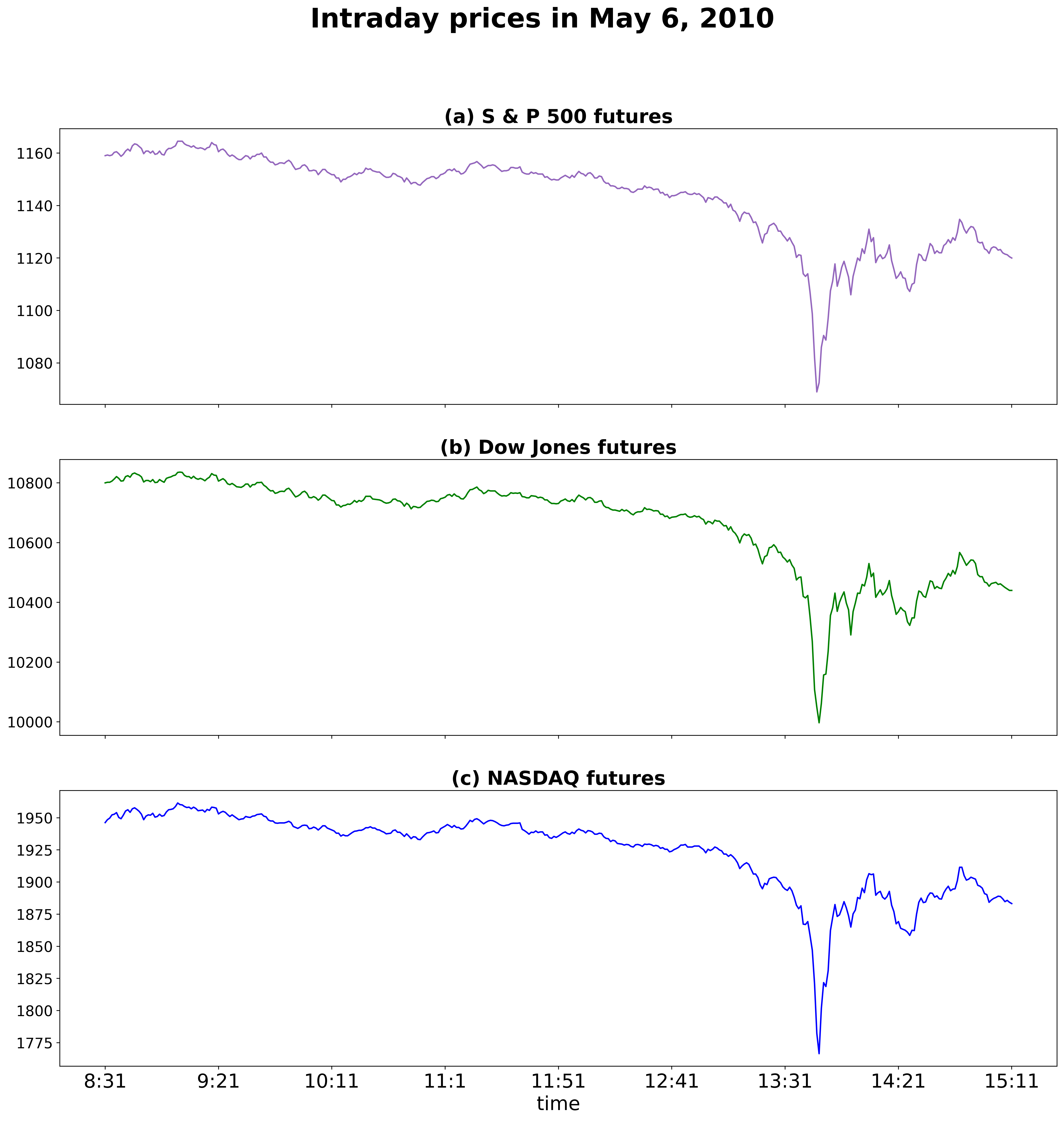}
\caption{Intraday price time series of S\&P 500 futures, Dow Jones futures, and NASDAQ futures on May 6, 2010.}
\label{fig1}
\end{figure}

In May 6, 2010, there was a sudden intraday stock market crash in U.S. financial markets known as \emph{Flash Crash}. The crash event, started at 2:32 p.m. EDT, lasted for approximately 36 minutes, during which major U.S. stock indexes such as the S\&P 500, Dow Jones Industrial Average, and Nasdaq Composite dropped larger than 6\% and rebounded very rapidly. Figure~\ref{fig1} shows intraday price time series of S\&P 500 futures, Dow Jones futures, and NASDAQ futures in May 6, 2010. Since such an extreme price-swing event lasting only for a very short time had never been reported before in financial markets, many financial researchers have been trying to understand the \emph{Flash Crash}~\cite{ref3,ref4,ref6,ref8}.

\section{Data and Methods} \label{sec:3}
We purchased one-minute price data for three futures based on three major U.S. stock market indexes: (a) S\&P 500 futures, (b) Dow Jones futures, and (c) NASDAQ futures from April 1, 2010 to May 28, 2010 (42 trading days) from BacktestMarket (\url{https://www.backtestmarket.com}). For each future $i$ ($i= a,b,c$) and for each minute $j$, we calculate one-minute log-return as $r_{i,j}=\log(\dfrac{P_{i,j}}{P_{i,j-1}})$, where $P_{i,j}$ represents the closing price of the futures $i$ at the minute $j$. Thus, for each minute $j$, we have a 3-dimensional vector defined by
\begin{equation*}
X_j=(r_{(a),j},r_{(b),j},r_{(c),j}).
\end{equation*}
Given a window size $w=50$, and for each $t\geq w$, we then construct the 3-dimesional time series defined by 
\begin{equation*}
\X_t=\{X_{t-49}, X_{t-48},\ldots,X_{t-1},X_t \}.
\end{equation*}
For each 3-dimensional time series $\X_t$, we compute $L^1$-norm of persistence landscape, $\lVert\lambda(\X_t )\rVert_1$, so we get a time series of $L^1$-norm of persistence landscape, 
\begin{equation*}
\Y=(Y_1,Y_2,\ldots,Y_n)=(\lVert\lambda(\X_w)\rVert_1, \lVert\lambda(\X_{w+1})\rVert_1,\ldots,\lVert\lambda(\X_{w+(n-1)})\rVert_1),
\end{equation*}
as in~\cite{ref5}. In order to detect an abnormality of the time series $\Y$, we develop a new abnormality measure based on the notions of Exponential Moving Average (EMA) and Exponential Moving Variance (EMVar) processes: Given the initial values,
\begin{equation*}
\mbox{EMA}_1=Y_1,\ \mbox{EMVar}_1=0,
\end{equation*}
the subsequent values, EMA$_i$  and EMVar$_i$, are computed using the following recursive formulae,
\begin{equation*}
\begin{array}{ccc}
\delta_i & = & Y_i-\mbox{EMA}_{i-1} \\
\mbox{EMA}_i & = & \mbox{EMA}_{i-1}+\alpha\cdot\delta_i \\
\mbox{EMVar}_i & = & (1-\alpha)\cdot (\mbox{EMVar}_{i-1}+\alpha\cdot\delta_i^2).
\end{array}
\end{equation*}
Then, our new abnormality measure $Z_t$ is defined by 
\begin{equation*}
Z_t=\dfrac{Y_t-\mbox{EMA}_{t-1}}{\sqrt{\mbox{EMVar}_{t-1}}}.
\end{equation*}
The new measure $Z_t$ represents the extent to which the current value $Y_t$ is deviated from the previous values.

\section{Empirical Results} \label{sec:4}
\begin{figure}[h]
\centering
\includegraphics[width=1.0\textwidth]{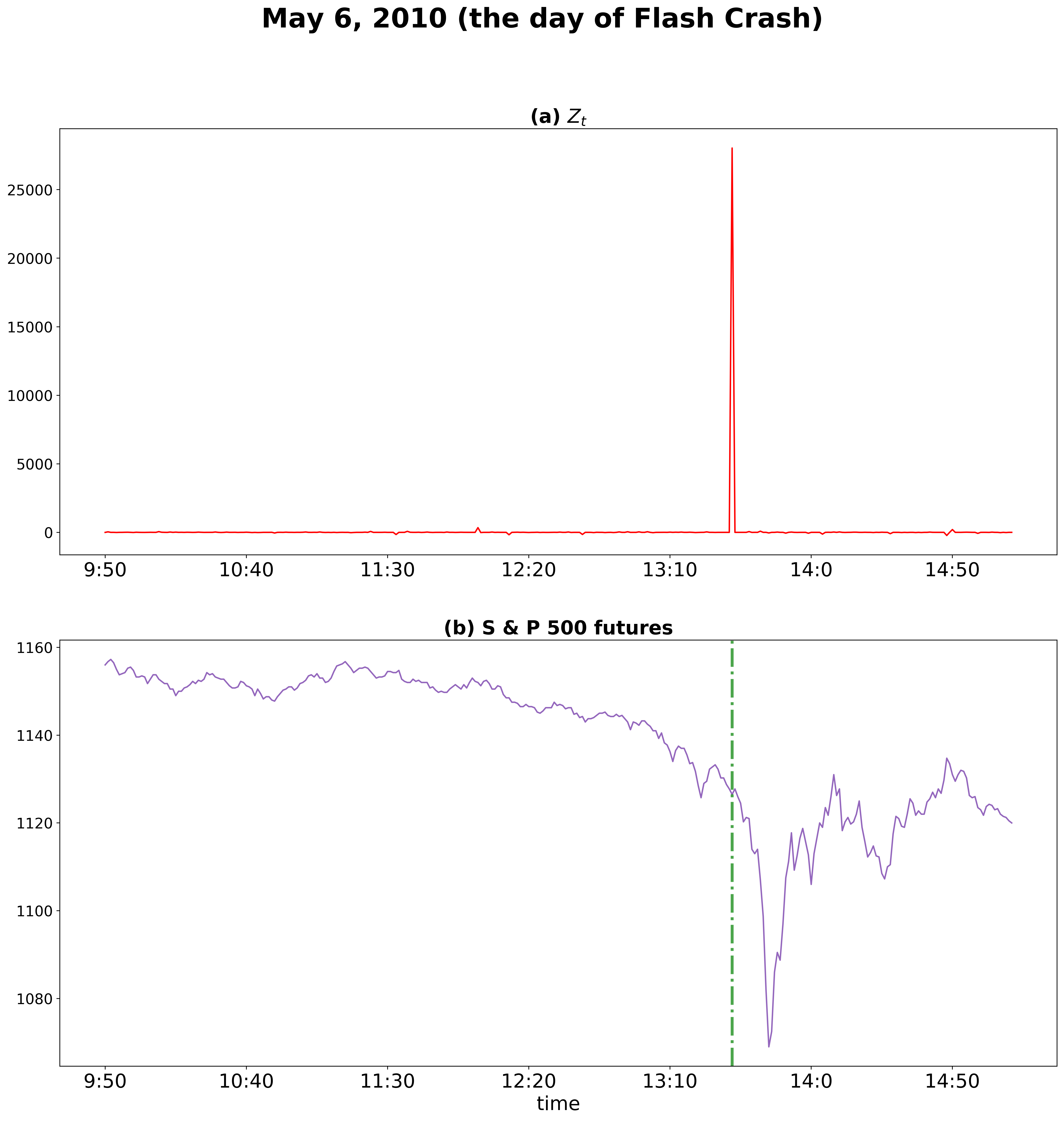}
\caption{Intraday time series plots of our new abnormality measure, $Z_t$, and prices of S\&P 500 futures on May 6, 2010, the day of Flash Crash.}
\label{fig2}
\end{figure}

\begin{figure}[h]
\centering
\includegraphics[width=1.0\textwidth]{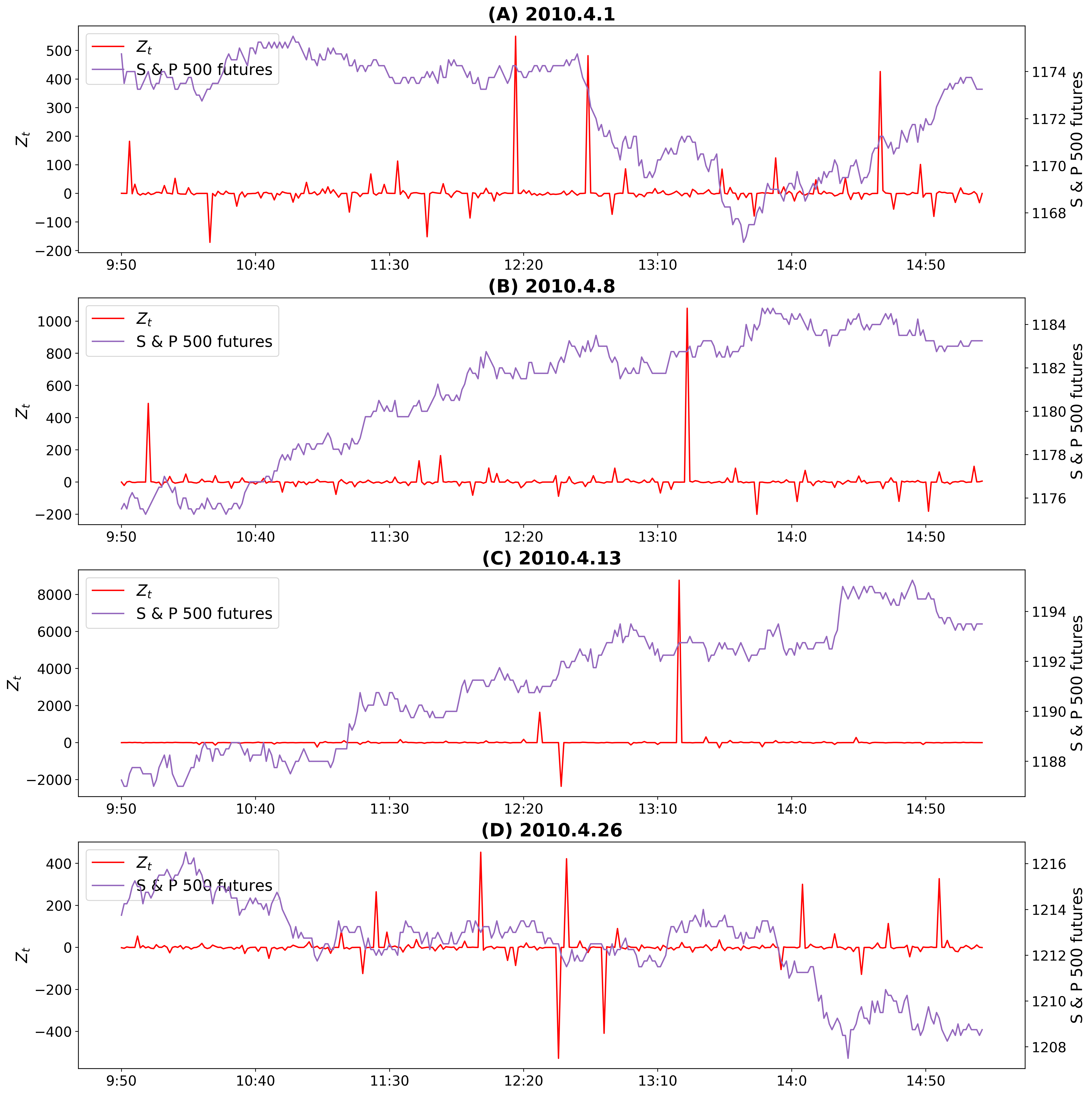}
\caption{Intraday time series plots of our new abnormality measure, $Z_t$, and prices of S\&P 500 futures on April 1, 8, 13, and 26, 2010.}
\label{fig3}
\end{figure}

\begin{figure}[h]
\centering
\includegraphics[width=1.0\textwidth]{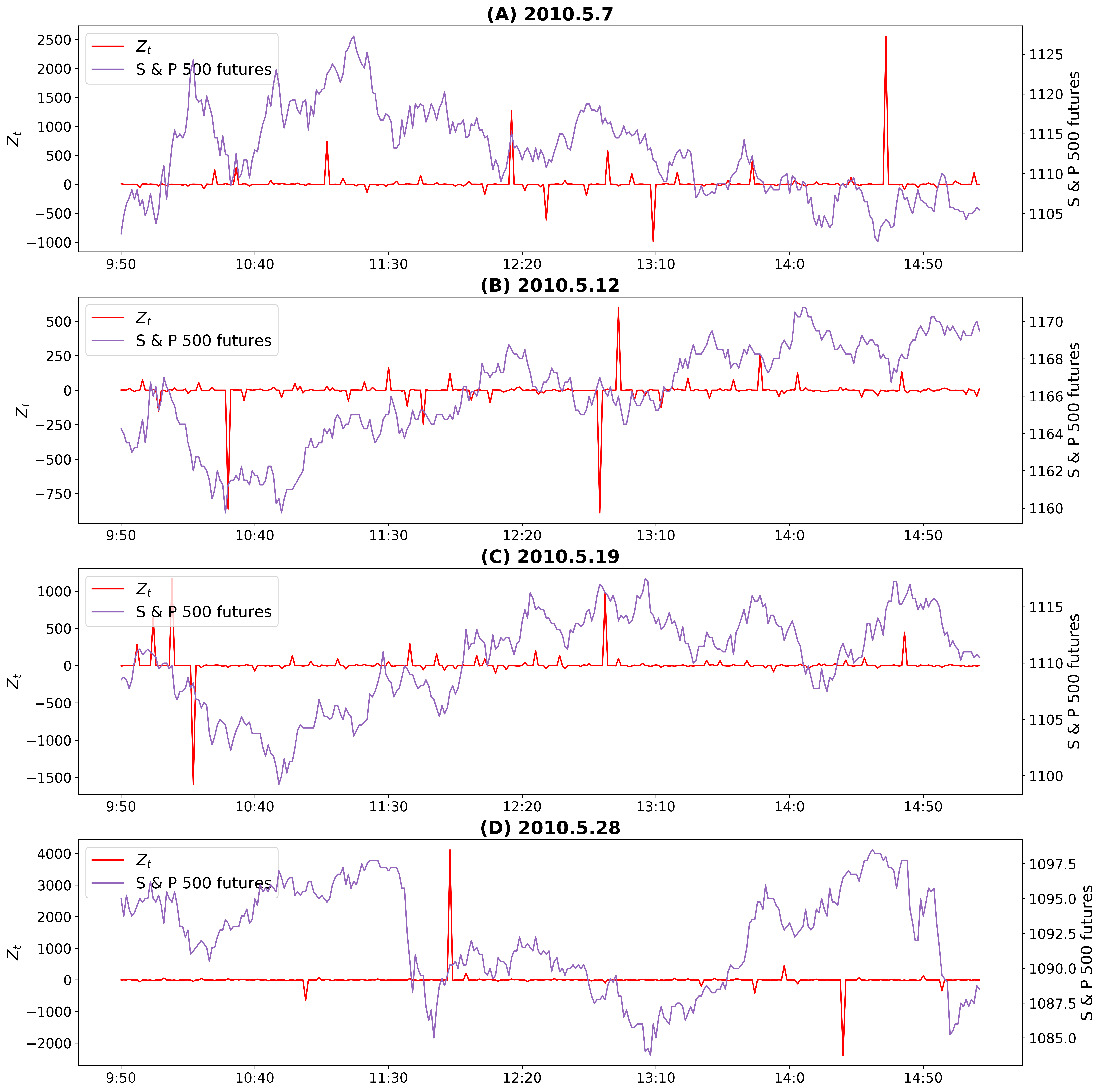}
\caption{Intraday time series plots of our new abnormality measure, $Z_t$, and prices of S\&P 500 futures on May 7, 12, 19, and 28, 2010.}
\label{fig4}
\end{figure}

Figure~\ref{fig2} displays the intraday time series plots of our new abnormality measure $Z_t$ and the prices of S\&P 500 futures on May 6, 2010, the day of Flash Crash. As Figure~\ref{fig2} shows, just before the extreme price-swing of Flash Crash begins, the abnormality measure $Z_t$ has the value of 28,034.92. The value is exceptionally huge compared to the values of $Z_t$ at other times of the day. Figure~\ref{fig3} (resp., Figure~\ref{fig4}) also presents the intraday time series plots of $Z_t$ and S\&P 500 futures in four randomly selected days before (resp., after) the day of Flash Crash out of our sample period. Figure~\ref{fig3} and~\ref{fig4} show that the largest value of $Z_t$ in Flash Crash day, 28,034.92, is also extremely large compared to the values of $Z_t$ in other days.

\section{Conclusion} \label{sec:5}
In this paper, we present a new method of detecting an abnormal phenomenon based on TDA and time series analysis, and investigate the event of Flash Crash via the new method. The empirical results from our study reveal that via the new method based on TDA, we can predict the event of Flash Crash. Therefore, our result shows that TDA not only can be used in forecasting long-term market crash events~\cite{ref5}, but also can be used in predicting intraday market crash events.

\end{document}